\documentclass[10pt,aps,twocolumn,showpacs,preprintnumbers,amsmath,amssymb,prb,superscriptaddress,footinbib]{revtex4-1}

\usepackage[utf8]{inputenc}
\usepackage{graphics}
\usepackage{latexsym}
\usepackage{amsmath}
\usepackage{float}
\usepackage{braket}
\usepackage{hyperref}

\newcommand{\PP}{\mathcal{P}}

\begin{document}

\title{Optimal frequency estimation and its application to quantum dots}
\author{\'Angel Guti\'errez-Rubio}
\affiliation{RIKEN Center for Emergent Matter Science, 2-1 Hirosawa, Wako, Saitama, 351-0198 Japan}
\author{Peter Stano}
\affiliation{RIKEN Center for Emergent Matter Science, 2-1 Hirosawa, Wako, Saitama, 351-0198 Japan}
\author{Daniel Loss}
\affiliation{RIKEN Center for Emergent Matter Science, 2-1 Hirosawa, Wako, Saitama, 351-0198 Japan}
\affiliation{Department of Physics, University of Basel, Klingelbergstrasse 82, CH-4056 Basel, Switzerland}
\date{\today}

\begin{abstract}
We address the interaction-time optimization for frequency estimation in a two-level system.
The goal is to estimate with maximum precision a stochastic perturbation.
Our approach is valid for any figure of merit used to define optimality, and is illustrated for the variance and entropy.
For the entropy, we clarify the connection to maximum-likelihood estimation.
We devise novel estimation protocols with and without feedback.
They outperform common protocols given in the literature.
We design a probabilistic self-consistent protocol as an optimal estimation without feedback.
It can improve current experimental techniques and boost coherence times in quantum computing.
\end{abstract}
\pacs{03.67.Ac, 89.70.Cf}
\maketitle

\section{Introduction}
\label{sec:introduction}

The task of measuring---or \emph{estimating}---the phase has a long history in physics.\cite{michelson1887on} In quantum mechanics, the existence of an operator $\hat{\phi}$ for the phase, conjugated to the number operator $\hat{N}$, has given rise to a rich body of literature.\cite{carruthers1968phase,nieto1993quantum,levy-leblond1976who}
One of the important outcomes of these studies is the notion of a \emph{canonical phase measurement} which is the optimal measurement of the phase operator that is allowed by the Heisenberg uncertainty principle.
Unfortunately, it is not known how to implement such a measurement in a realistic setting.\cite{wiseman1995adaptive,martin2020implementation}

The situation is different for a classical phase $\phi$ which is a well-defined parameter without quantum fluctuations.
Such an approach is standard in optical interferometry, where $\phi$ enters in a so-called phase-shift operator, $\exp(i \phi \, b^\dagger b)$.
Here, $b$ is a destruction operator of a photon passing through one arm of the interferometer.
A typical optimization task can then be formulated like this: Given $N$ photons, how do we send them through the interferometer in order to learn as much as possible about the value of $\phi$?
In other words, how do we optimize over many-body photon states?
While the formal solution to this task is known,\cite{berry2000optimal,knysh2014true} being related to the canonical measurement operator, it is of little practical value since such states are hardly realizable in the laboratory.
Thus, the practicality of optimization procedures becomes a key aspect in estimation.\cite{berry2000optimal}

There are useful guidelines for judging how good an estimator is.
In the above scenario with a fixed number of photons, the Heisenberg uncertainty relations imply that the variance estimator\footnote{We abuse the notation by not discriminating the value of the phase $\phi$ and its estimator, being a function of measured results, sometimes written as $\phi_E$ or $\hat{\phi}$.} $\textrm{var} (\phi)$ can not be smaller than $1/\mathcal{N}^2$.\cite{ou1997fundamental} In other words, quantum mechanics sets the \emph{Heisenberg limit} (or Heisenberg scaling) $\textrm{var} (\phi) \sim 1/\mathcal{N}^2$.
Importantly, classical photon states (such as the coherent light of a laser) or other simple states (such as unentangled single photons) perform worse.\cite{caves1980quantum-mechanical} Their scaling $\textrm{var} (\phi) \sim 1/\mathcal{N}$ is called the \emph{standard quantum limit} (SQL). 
Overcoming the SQL using squeezed light,\cite{caves1981quantum-mechanical} meaning using quantum resources, is an exemplary achievement.\cite{goda2008quantum-enhanced,the-ligo-scientific-collaboration2011gravitational}

The estimator formulation using an operator has been generalized in quantum information theory to the \emph{quantum phase estimation} (QPE) problem.~\cite{cleve1998quantum,obrien2019quantum} Here the task is to estimate an eigenvalue of a unitary (in general, many-body) operator $U$.
Interestingly, all known quantum algorithms can be recast as the QPE problem.\cite{cleve1998quantum} Nevertheless, to stay close to the previous discussion and to the main topic of this article, let us consider a single-qubit operator $U=\exp( i \phi \sigma_z)$.
Here, $\sigma_z$ is a Pauli matrix.
In QPE, one is allowed to insert an arbitrary number of applications of $U$ into a general quantum circuit.
The \emph{quantum phase estimation algorithm} (QPEA)\cite{cleve1998quantum} and the \emph{Kitaev algorithm}\cite{kitaev1997quantum} are two closely related solutions to the QPE problem.
Both algorithms invoke circuits with an exponential set of powers $U^n$, that is $n \in \{ 2^k \mid 0 \leq k \leq K\}$.\footnote{Ref.~\onlinecite{svore2013faster} investigates a variant of QPEA including additional values of $n$.} The idea is that the power $n=2^k$ probes the $k$-th bit of $\phi$.
The discovery in Ref.~\onlinecite{griffiths1996semiclassical} allows one to implement QPEA\footnote{From now on, we shorten ``QPEA and Kitaev algorithm'' into just ``QPEA''.} by circuits without the need of entangling gates, trading them instead for single-qubit feedback gates.

The problem with QPEA is that the required quantum circuits, with error-free measurements and gates (including $U$), do not exist in practice so far.
In the current, \emph{noisy intermediate-scale quantum} (NISQ) hardware the noise is not small: While the error-free QPEA is optimal in a certain sense,\footnote{While QPEA does not reach the Heisenberg scaling,\cite{berry2009how} it learns one bit of $\phi$ from one bit of measurement result, which is as much as possible with an overhead of order one and an overall probability of success of order one (all bits estimated correctly).
}  it will fail on current state-of-the-art hardware.\cite{mcclean2014exploiting}
Moreover, the properties of the noise depend on the specific quantum circuit: On the physical platform, the device, quantum gates, measurements, and so on.
This dependence makes the \emph{optimal} estimation hinge on particular details, and thus hard to formulate in generic terms.
We believe that in this case it is more appropriate to describe $U$ as being dynamically generated by a time-dependent Hamiltonian $H$ which is turned on during a specific interaction time $t$.
Compared to interspersing the circuit with noise gates, such a dynamical description is more faithful\cite{rispler2020towards} and also more amenable to  performance comparisons across platforms.
For the above single-qubit example, the corresponding Hamiltonian is $H = \hbar \omega \sigma_z$.
The noise enters here naturally as a limited control over the value and stability of parameters such as the frequency $\omega$.
In other words, these parameters fluctuate, both in time and in other variables such as space or devices.
In Sec.~\ref{sec:discussion} we discuss additional reasons why the specific nature of the noise has to be a part of the problem formulation.

The present work is about the optimal estimation of a fluctuating frequency $\omega$ in the above Hamiltonian which occurs in the context of typical single-qubit experiments.
In such experiments, the qubit is repeatedly (indexed by integer $n$) initialized to state $(|0\rangle + |1\rangle)/\surd{2}$, evolved under $H$ during variable interaction time $t$, and projectively measured in the basis $\{ |0\rangle, |1\rangle \}$.
What is being optimized is the set of interaction times $\{\tau_1, \dots, \tau_N\}$, a set of $N$ positive real numbers.
\footnote{To keep the correspondence to interferometry with photons, the resource counted is the total interaction time $\mathcal{N} \equiv  (\sum_n \tau_n)/\tau_1$, with $\tau_1$ a time unit converting one photon to the interaction time.
Counting the number of measurements, $\mathcal{N} \equiv N$ might be more appropriate if dead times dominate (the interaction time is only a small part of the whole measurement cycle duration),\cite{sergeevich2011characterization} which is sometimes the case in solid-state qubit experiments.}
The motivation for this formulation has both practical and theoretical reasons.
As to the first reason,\cite{huszar2012adaptive} instead of relying on entangled many-qubit states or many-qubit measurements,\cite{nagata2007beating} the above procedure is realizable on current NISQ hardware.
As to the second reason, it has been found that this procedure is enough to reach Heisenberg scaling with photons.\cite{berry2009how} Additionally, this task is at the core of magnetometry using ions, \cite{ruster2017entanglement-based} superconducting qubits \cite{danilin2018quantum-enhanced} or NV centers, \cite{bonato2016optimized} or of precise time measurements.\cite{mullan2014optimizing}
Finally, we are also motivated by current experiments with solid-state qubits, where better and faster estimation extends not only the qubit coherence \cite{shulman2014suppressing,delbecq2016quantum} but also increases the fidelity of gates\cite{noiri2018fast} and measurements.\cite{nakajima2019quantum, yoneda2020quantum}

Putting aside idealized or non-implementable solutions, the above optimization task even for a single qubit does not have an analytical solution.
Indeed, a formidable problem arises here: If $N$ is large, a brute-force optimization is impossible, as the configuration space is exponentially large in $N$.
Still, there are two main approaches left to proceed: In the first one, the \emph{global} optimization (of the estimator after $N$ measurements) is traded  for $N$ consecutive \emph{local} optimizations.
This means that the interaction time $\tau_{n}$ is chosen as the one that maximizes the expected value of the figure of merit\cite{FootnoteFOM} in the $n$-th measurement.
However, it is not known how well such a local optimization fares compared to a global one.\cite{higgins2009mixed,berry2001optimal} Second, the local maximization requires taking into account all previous measurements up to the $(n-1)$th.
We call this procedure \emph{estimation with feedback}.

Unfortunately, implementing feedback is often technically too demanding.
In that case one resorts to \emph{estimation without feedback}, and instead prescribes the set of interaction times before the experiment starts.
So far, only \emph{heuristic} optimizations for  such tasks without feedback have been given in the literature.
By this we mean the following: One guesses the overall structure of the set $\{\tau_1, \dots, \tau_N\}$ in some way, parametrizing it by a few numbers, and then optimizes over these few remaining degrees of freedom.
Typical examples of such heuristic guesses are: (i)  constant interaction times $\tau_n = \alpha$, (ii) linearly growing ones,\cite{obrien2019quantum,sergeevich2011characterization,shulman2014suppressing}  $\tau_n = \alpha + (n-1) \beta$, or (iii) exponentially distributed ones,\cite{danilin2018quantum-enhanced,bonato2017adaptive,bonato2016optimized,hayes2014swarm,said2011nanoscale,berry2009how,higgins2009demonstrating,higgins2007entanglement-free} $\tau_n = \alpha 2^{n-1}$.
The third scenario is inspired by  QPEA and uses interaction times with a power $k$ of multiplicity $N = F + G(K-k)$ in the set.
Here, the parameters $\alpha$, $\beta$, and $F$, $G$, and possibly $K$, are being optimized, respectively.
Local and heuristic optimization procedures can therefore be seen as two different classes of tractable algorithms.
Both are based on a drastic restriction of the space within which the optimal interaction times are searched for.

The main achievement of this work is to propose a new class of optimal estimation protocols which are without feedback but nevertheless are free of heuristic guesses.
Therefore, the proposed protocol searches the solution within a vastly bigger space, and yet is both optimal (within a given class, precisely defined below) and numerically feasible (its construction is $O(N^2)$ at most, but based on the example we study we conjecture it will be $O(N)$ in many practical cases).
The main idea behind our protocol is that we parametrize each interaction time $\tau_n$ by a probability distribution, which is not restricted in advance in any way.
The interaction times $\tau_n$ are therefore stochastic variables, generated probabilistically.
Their probability distributions are nevertheless unique, explicitly constructed by a numerical algorithm provided below.
They depend on the system details such as the  Hamiltonian, the measurement, the estimated variable dynamics, and others.
Finally, the construction we propose is robust with respect to noise, both in the circuit and in the estimated variable, to the variable prior probability distribution, and naturally encompasses both \emph{localization} and \emph{tracking} regimes (see Sec.~\ref{sec:noise} for their definitions).

The article is organized as follows.
In Sec.~\ref{sec:system}, we describe the system and the procedure of Bayesian estimation.
In Sec.~\ref{sec:mle}, we introduce our figures of merit and the relation to maximum likelihood.
In Sec.~\ref{sec:optimal}, we design the estimation protocols and evaluate them with numerical simulations.
Sec.~\ref{sec:discussion} contains an extension of the discussion started in the introduction and the connection of our work to the existing results in the literature.
There we focus on aspects which are easier to understand after the details of our method have been explained and demonstrated.
In Sec.~\ref{sec:conclusions}, we summarize our results and conclude.

\section{Bayesian estimation on a two-level system}
\label{sec:system}

Let $H(t) = \frac{\hbar}{2} [\Delta + \Gamma(t)] \sigma_x$ be the Hamiltonian of a two-level system.
Define $\Omega(t) = |\Delta + \Gamma(t)|$  as the frequency of the oscillations between the eigenvectors of $\sigma_z$, denoted  by $\ket{\downarrow}$ and $\ket{\uparrow}$; let $\Delta > 0$ be constant, $\Gamma(t)$ a time-dependent perturbation and assume $\langle \Gamma(t) \rangle = 0$, $|\Gamma(t)|<|\Delta|$.
$\Gamma(t)$ is the source of dephasing with respect to the average frequency $\Delta = \langle \Omega(t) \rangle$.

This system is particularly interesting because its Bayesian update (defined below) has a closed solution for all diffusion regimes.
App.~\ref{app:hamiltonian} also discusses a general two-level Hamiltonian and to what extent the results extrapolate.
Moreover, our Hamiltonian maps to a prospective qubit, a two-electron double quantum dot in $\rm{GaAs}$.
In that case, $\ket{\downarrow}$ and $\ket{\uparrow}$ represent the singlet and triplet states, $\Delta$ is a micromagnet gradient and $\Gamma(t)$ is the difference between the projections of the Overhauser field on the micromagnet field.\cite{delbecq2016quantum,shulman2014suppressing,sergeevich2011characterization}

Our goal is to detect and measure in time the value of $\Omega(t)$ with as much precision as possible.
Not only for the sake of its measurement, but also to mitigate dephasing.
With \emph{estimation} of a variable, we mean giving its probability distribution.
To measure $\Omega(t)$, we estimate $\Omega(t_i)$ at a discrete set of times $t_i$, $i = 1, \dots, N$.
These are called \emph{initialization times} in the typical experiment described below.

To simplify the notation, $\omega$ denotes $\Omega(t_i)$ irrespective of the index $i$.
The context that will accompany $\omega$ will make this notation unambiguous.
Uppercase symbols like $\Gamma$ or $\Omega$ represent functions of time and usually omit the time argument.

The estimation of $\omega$ involves two steps: The Bayesian update after each qubit measurement and the diffusion of $\omega$ from then on.
In this section, we separately outline both.
After that, we explain how to apply them sequentially and repeatedly in a typical experiment.
This is all we need to design estimation protocols.
We note in passing that alternating the update and diffusion is a standard approach, for example in recursive Bayesian filtering.\cite{doucet2001sequential}

Qubit measurements are the only way to access $\Omega$.
To describe the Bayesian update, let us look first at a single projective measurement.
Assume that the system is prepared in the initial state $\ket{\uparrow}$ at time $t_i$, and let $\omega = \Omega(t_i)$ be the unknown frequency at initialization.
After the evolution during time $\tau$, the probability that a measurement yields $\ket{\downarrow}$ ($m = -1$) or $\ket{\uparrow}$ ($m = +1$) reads
\footnote{
This expression is straightforward from elementary probability rules:
\begin{align*}
    P & (m | \omega, \tau)
    =   
    P(m, \smallint \mathcal{D}\bar{\Omega} \,
         \bar{\Omega} | \omega, \tau)
    = \int \mathcal{D}\bar{\Omega} \,
        P(m, \bar{\Omega} | \omega, \tau)
    \\
    & = \int \mathcal{D}\bar{\Omega} \,
        P(\bar{\Omega} | \omega, \tau)
        P(m | \bar{\Omega}, \omega, \tau)
    \\
    & = \int \mathcal{D}\bar{\Omega} \,
        P(\bar{\Omega} | \omega)
        P(m | \bar{\Omega}, \tau)
    = \int \mathcal{D}\Omega \,
        P(\Omega) P(m | \Omega, \tau) \,.
\end{align*}
Here, the functional integral in $\bar{\Omega}$ runs over all arbitrary functions of $t$; and the integral in $\Omega$, as we defined in the main text, over all functions with $\Omega(t_i) = \omega$.
Throughout this article, we follow the propositional notation common in probability theory:\cite{jaynes2003probability} arguments inside $P$ represent logical statements.
Accordingly, sums or integrals act as the logical `or' operator.
}
\begin{align}
    P(m | \omega, \tau)
    = 
    \int \mathcal{D}\Omega \,
    P(\Omega)
    P(m | \Omega, \tau)
    \,,
    \label{eq:Pmw}
\end{align}
with 
$
    P(m | \Omega, \tau) = |\bra{m}
    \exp[-\frac{i}{\hbar} \int_{t_i}^{t_i+\tau}dt \, H(t)]
    \ket{\uparrow}|^2
$.
The functional integral in Eq.~\eqref{eq:Pmw} runs over all possible evolutions $\Omega(t)$, with weighting probability $P(\Omega)$---derived in App.~\ref{app:proof_closed}.
Moreover,
\begin{align}
    P (m | \Omega, \tau)
    = \frac{1}{2}
    \left(
        1 + m\cos\int_{t_i}^{t_i+\tau} dt\, \Omega(t)
    \right) \,.
    \label{eq:Pmt}
\end{align}

We emphasize that our approach, outlined by Eqs.~\eqref{eq:Pmw} and \eqref{eq:Pmt}, is valid for arbitrary changes of $\Omega(t)$ during the time $\tau$.
In contrast, the literature\cite{shulman2014suppressing,delbecq2016quantum,dinani2019bayesian,bonato2017adaptive} often approximates $\Omega(t) \simeq \Omega(t_i)$ for $t \in [t_i, t_i+\tau]$, and obtains $P(m | \omega, \tau) \simeq [1+m\cos(\omega \tau)]/2$.
This approximation only holds for slow diffusion of $\Omega(t)$,
\footnote{
As an example, in Ref.~\onlinecite{bonato2017adaptive} one reads that `an exact calculation \dots could potentially result in more accurate estimates, but the method to perform such a calculation appears to be an open question.' We answer that question here, among others.
}
as we discuss after Eq.~\eqref{eq:closed}.

Knowing $P(m | \omega,\tau)$, we can apply Bayes' rule to estimate $\omega$.
Bayes' rule updates the prior distribution $P(\omega)$ to the so-called posterior distribution $P(\omega | m, \tau)$ after the interaction time $\tau$ and the outcome $m = \pm 1$,
\begin{align}
    P(\omega | m, \tau)
    = \frac{P(m | \omega, \tau) P(\omega)}{P(m | \tau)} \,.
    \label{eq:Bayes}
\end{align}
$P(m | \tau)$ can be calculated as a normalization constant.
We emphasize that in these expressions, $\omega$ represents the value of $\Omega(t)$ at the \emph{last} initialization time.

Now, we analyze how the diffusion of $\Omega(t)$ affects frequency estimation.
It already entered the problem through the functional integral of Eq.~\eqref{eq:Pmw}.
But it also broadens $P(\omega)$ onto $P^d(\omega)$ after a time $\delta t$,
\begin{align}
    \label{eq:diffusion}
    P^d(\omega)
    = \int_{-\infty}^{\infty} d\omega' \,
    P(\omega')
    K(\omega', \omega, \delta t) \, .
\end{align}
$K(\omega', \omega, \delta t)$ is the diffusion kernel that encodes the time dynamics of $\Omega(t)$.
To keep the demonstration of our construction of optimal estimation protocols reasonably simple, we assume that the kernel is Markovian.
Were it not the case, one would have to implement, most probably numerically, a way to obtain $P^d(\omega)$ from $P(\omega)$, and then proceed as we describe below.
Concerning also other generalizations, see the discussion in App.~\ref{app:hamiltonian}.

A model for diffusion completes the description of the system.
Our choice is the random walk in the presence of a harmonic potential,\cite{uhlenbeck1930on,rabenstein2004qubit} as a minimal model describing fluctuations that occur around zero with typical deviation $\sigma_{\Omega}$ in a timescale given by $\kappa$.
On the one hand, one can consider this as a two-parameter approximation to any Markovian fluctuation, encoding its range and speed.
On the other hand, a simple bounded-diffusion model is adequate to describe the dynamics of a large ensemble of thermally disordered nuclear spins in a quantum dot,\cite{deng2005nuclear} being our practical example adopted for an illustration.
Using differences from the mean, $\delta \omega = \omega - \Delta$ and $\delta \omega^\prime = \omega^\prime - \Delta$, the diffusion kernel for this model is
\begin{align}
    K(\omega^\prime, \omega, \delta t)
        = \frac{1}{\sqrt{2\pi\sigma_{\delta t}^2}}
    \exp
    \left[
        -\frac{(\delta \omega-\delta \omega^\prime
               e^{-\delta t/\kappa})^2}
              {2\sigma_{\delta t}^2}
    \right] \,,
    \label{eq:kernel}
\end{align}
where
$
    \sigma_{\delta t}^2
    = \sigma_\Omega^2 (1-e^{-2\delta t/\kappa}) \,.
$

The kernel $K(\omega', \omega, \delta t)$ gives the probability that $\omega'$ evolves to $\omega$ after a time $\delta t$.
Therefore, it determines $P(\Omega)$ in Eq.~\eqref{eq:Pmw}.
As App.~\ref{app:proof_closed} shows, Eq.~\eqref{eq:kernel} allows us to express Eq.~\eqref{eq:Pmw} in closed form,
\begin{equation}
    P(m | \omega, \tau)
    = \frac{1}{2}
    \left[
        1 + m \, e^{-\psi(\tau)}
        \cos  \phi(\omega, \tau)
    \right] \,.
    \label{eq:closed}
\end{equation}
Here,
\begin{align}
    \psi(\tau)
    & \equiv 
    \sigma_\Omega^2 \kappa
    \left[
        \tau + \frac{\kappa}{2}
        (1-e^{-\tau/\kappa})(-3+e^{-\tau/\kappa})
    \right]
    \,, \notag \\
    \phi(\omega, \tau)
    & \equiv
    \tau \Delta + \kappa (\omega - \Delta)
    (1-e^{-\tau/\kappa}) \,.
    \notag
\end{align}
The first-order expansion in $\tau/\kappa$ reduces Eq.~\eqref{eq:closed} to an expression with $\psi(\tau) \simeq 0$ and $\phi(\omega, \tau) \simeq \omega \tau$.
This is the approximation of slow diffusion mentioned below Eq.~\eqref{eq:Pmt}.
For what follows, it is essential that we do not adopt this approximation and we use Eq.~\eqref{eq:closed}.

\begin{figure}
    \centering
    \includegraphics{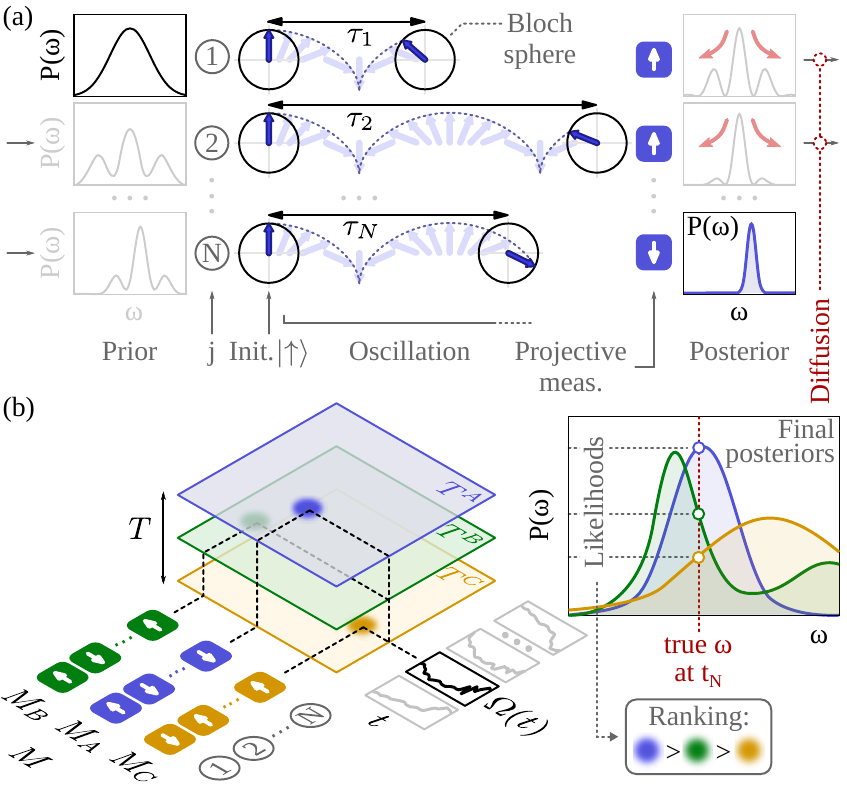}
    \caption{(a) Outline of a frequency-estimation experiment for a given protocol $T = \{\tau_1, \dots, \tau_N\}$.
    The distributions of $\omega$ prior (posterior) to each measurement appear to the left (right) of the oscillation pictures.
    The red arrows represent diffusion.
    (b) The left side depicts the configuration space $(\Omega, T, M)$ and three experiments (colored dots).
    The protocols $T^A$, $T^B$, and $T^C$ applied to the same trajectory $\Omega(t)$ result in the different measurements $M_A$, $M_B$, and $M_C$, respectively.
    The final posterior distribution of each experiment appears on the right in the same color.
    The knowledge of the true final frequency $\Omega(t_N)$ would clearly rank the performance of the protocols.
    This is the basis of our discussion on the figures of merit and maximum likelihood.
}
    \label{fig:configuration_space}
\end{figure}

Bayesian update and diffusion are applied alternatingly and repeatedly.
They compete in the estimation of $\Omega$, roughly narrowing and smearing the probability distribution $P(\omega)$, respectively.
We lay out a typical experiment and pinpoint the changes in $P(\omega)$, see also Fig.~\ref{fig:configuration_space}a.
First, at time $t_1$ we initialize the system in $\ket{\uparrow}$, and after oscillating for a time $\tau_1$ of our choice, we perform a projective measurement.
Using the result $m_1$, Eq.~\eqref{eq:Bayes} and Eq.~\eqref{eq:closed} narrow the probability distribution of the frequency.
Eq.~\eqref{eq:diffusion} diffuses the resulting posterior until a new measurement is performed.
We repeat the whole procedure $N$ times, and after a set of measurements $M \equiv \{m_1, \dots, m_N\}$ corresponding to the interaction times $T \equiv \{\tau_1, \dots, \tau_N\}$, one obtains the posterior probability distribution $P(\omega | TM)$.
As we pointed out before, here $\omega$ corresponds to $\Omega(t_N)$, with $t_N$ the initialization time of the $N$th measurement.
For the prior $P(\omega)$ at the beginning of the experiment, we take a Gaussian centered at $\Delta$ with a dispersion $\sigma_\Omega$.
This is the least informative prior one can take given the mean value and variance of $\Omega$,\cite{jaynes1957information} and is an excellent approximation for the Overhauser field.\cite{merkulov2002electron}

We conclude this section by describing the configuration space of the problem.
It contains the main variables and helps to understand their relations.
Every experiment is uniquely represented by the coordinates $(\Omega, T, M)$.
$\Omega$ and $M$ reflect two different sources of randomness: One is the stochastic evolution of $\omega$; the other comes from the quantum probabilistic nature of projective measurement (so called quantum noise).
In contrast, the interaction times $T = \{\tau_1, \dots, \tau_N\}$ are up to our choice.
It is within this freedom where we consider optimality.
A procedure to choose $T$, and $T$ itself, will be called a \emph{protocol}.

Now that we have described the system, we can rephrase our main goal more precisely: We want to find $T$ for the optimal estimation of $\Omega$.
It only remains to properly define \emph{optimality}.
To do so, we discuss different figures of merit and their relations in the next section.

\section{Variance, entropy, and maximum-likelihood estimation}
\label{sec:mle}
In the literature on estimation, terms like \emph{optimal} or \emph{accuracy} are 
associated with different quantities that are not necessarily equivalent.\cite{bode1950simplified,berry2001optimal}

On the one hand, the variance is ubiquitous in quantum measurement.
It quantifies the error and defines the shot-noise and Heisenberg limits.\cite{ou1997fundamental,higgins2009demonstrating}
Some modifications like the Holevo variance are used in the estimation of periodic quantities like phases.\cite{wiseman1997adaptive,cappellaro2012spin-bath,berry2000optimal}

On the other hand, the entropy takes over in the context of information theory due to the Shannon theorem.
Adopting this figure of merit is extraordinarily far-reaching.
For example, one can build statistical mechanics almost exclusively from entropy maximization.\cite{jaynes1957information,goold2016role}
The fundamental binomial and Gaussian distributions also maximize the entropy.\cite{jaynes2003probability}

It is possible to justify the use of the entropy by arguments other than the relation to \emph{knowledge} or \emph{information}.
In the following we show how it emerges naturally from maximum-likelihood estimation.
For a protocol $T$, let $L_T$ be the log likelihood of estimating the true frequency at $t = t_N$.
And let $\langle S_T \rangle$ be the mean entropy of the final posteriors $P(\omega | TM)$ after sampling the whole $(\Omega, M)$ subspace.
This subspace is a horizontal plane in Fig.~\ref{fig:configuration_space}b (left).
App.~\ref{app:proof} proves the relation:
\begin{align}
    L_T \equiv
    \sum_{\Omega, M}
    P(\Omega M | T)
    \log P \boldsymbol{(}
        \omega = \Omega(t_N) | TM
    \boldsymbol{)}
    = -\langle S_T \rangle \,.
    \label{eq:theorem}
\end{align}
In other words, optimizing the entropy also optimizes the maximum average likelihood.
The crucial difference is that the evaluation of the right hand side does not require to know the true frequency, whereas the left hand side does.
In sum, Eq.~\eqref{eq:theorem} connects maximum-likelihood estimation with entropy optimization.

To give an intuitive understanding of Eq.~\eqref{eq:theorem}, let us illustrate with 
Fig.~\ref{fig:configuration_space}b showing the case of a single trajectory $\Omega(t)$ and three protocols $T^A$, $T^B$, and $T^C$.
The ranking `$T^A$ better than $T^B$ and $T^B$ better than $T^C$' is straightforward from the plot on the right.
As we prove in App.~\ref{app:proof}, the left-hand side of Eq.~\eqref{eq:theorem} is the cumulative application of the same criterion for all possible experiments given a protocol $T$.
Hence the sum over $\Omega$ and $M$.
On the right-hand side, the resulting quantity is the mean entropy of the final distributions $P(\omega | TM)$.

A given variance sets an upper bound for the entropy,\cite{berry2001optimal} but the converse is not true.
Consequently, the variance and the entropy are related but not equivalent.
This raises a fundamental question in parameter estimation: What is preferable, to optimize the variance and get an estimation with the least squared error, or to minimize the entropy and thus maximize the likelihood of guessing the variable right?
Actually, there is no categorical answer: Rejecting either sacrifices optimality in a different manner.
One must choose the most suitable figure of merit for a specific purpose.

Nevertheless, the protocol's likelihood $L_T$, given by $\langle S_T \rangle$, is a natural benchmark for protocols in our configuration space.
$\langle S_T \rangle$ stands out versus other in principle valid quantities: For example, instead of considering the average of the entropy $\langle S_T \rangle$, why not take the entropy of the averaged error distributions?
With our derivation, we can answer that $\langle S_T \rangle$ has a deeper meaning in terms of likelihood.
For this reason, we focus on $\langle S_T \rangle$ to compare protocols from now on, but also discuss the variance for the sake of completeness.

\section{Optimal frequency estimation}
\label{sec:optimal}

In this section, we design protocols that can optimize an arbitrary figure of merit.
We illustrate them for the entropy $\langle S_T \rangle$ and the variance.
$T^S$ and $T^\sigma$ denote these protocols, respectively.
We design these protocols with and without feedback.
By feedback, we mean using the outcomes $M_{j-1} \equiv \{ m_1, \dots, m_{j-1} \}$ to set $\tau_j$.
This is denoted by an asterisk as in $T^{S*}$.
On the contrary, protocols without feedback set all $\tau_j$ before the experiment starts.
Some works call these protocols \emph{online} and \emph{offline}, respectively.

We limit our protocols to a \emph{memoryless} choice of interaction times: $\tau_{k}$ for $k < j$ does not directly influence $\tau_j$, but only indirectly through Bayesian update (explained in detail below).
The literature sometimes refers to this as \emph{local} optimization.
Proceeding otherwise (or \emph{globally}), one would have to sample a space of exponentially growing dimension and thus apply Monte Carlo techniques.\cite{doucet2000on}
This study is outside the scope of this article and we defer it to future works.

By numerical simulations, we compare our protocols to others inspired by the literature.
First, we compare them with a set of linear protocols\cite{obrien2019quantum,shulman2014suppressing,sergeevich2011characterization,delbecq2016quantum,ruster2017entanglement-based} $\{T^{\rm{lin},k} \, | \, k = 1,\dots,K\}$, each of them with $\tau_j = \alpha_k j$, $j = 1,\dots,N$ and $\alpha_k$ constant.
Second, with the \emph{saw-toothed} protocol $T^{\rm{saw}}$ with $\tau_j = \alpha_m \times (j \mod n)$ for some $m \leq K$ and $n < N$.

Our simulations take as a benchmark the double quantum dot from Refs.~\onlinecite{delbecq2016quantum,shulman2014suppressing} mentioned in Sec.~\ref{sec:system}.
But we make one important modification: We deal with fast diffusion, $\delta t/\kappa \sim 1$.
The reason is that this makes our numerical demonstration more general.
Namely, fast diffusion can be trivially extended to slow diffusion, but not the other way around.
Consequently, we take $\kappa \simeq 3.1\times 10^{-2}\, \rm{s}$, three orders of magnitude smaller than the typical value for the Overhauser field.\cite{paget1982optical}
For the rest of the parameters, we adopt $\sigma_\Omega = 10\ \rm{MHz}$, $\Delta = 30\ \rm{MHz}$, $N = 150$ and $\delta t \simeq 15\, \rm{ms}$.
1000 simulated experiments are run for each protocol $T$ to sample the configuration space.

\subsection{Protocols with feedback}
\label{sec:feedback}

In this section, we design a protocol with feedback $T^{X*}$ to optimize an arbitrary figure of merit $X$.
We illustrate it for the entropy, and we show the results also for the variance.

The protocols with feedback set $\tau_j$ as follows.
Before measurement $j\leq N$, we have the past outcomes $M_{j-1} = \{m_1, \dots, m_{j-1}\}$ for the interaction times $T_{j-1}^{S*} \equiv \{\tau_1^*, \dots, \tau_{j-1}^*\}$, and thus $P(\omega | T_{j-1}^{S*} M_{j-1})$.
We choose the optimal interaction time $\tau_j^*$ that minimizes the expected value of the entropy right before measurement $j+1$,
\begin{align}
    \bar{S}_j 
    = \sum_{m_j = \pm 1}
    P(m_j | T_j^{S*} M_{j-1})
    S[P^d(\omega | T_j^{S*} M_j)] \,.
    \label{eq:S_j}
\end{align}
Here, the first factor can be calculated as
$
    \int_{-\infty}^{\infty}
    d\omega \, 
    P(m_j | \omega, \tau_j) 
    P^d(\omega | T_{j-1}^{S*} M_{j-1})
$.
$S[P(\omega)]$ is the entropy of the probability distribution $P(\omega)$.
In this way, by minimizing $\bar{S}_j$ at each measurement $j$, we aim at reducing $\langle S_T \rangle$ as much as possible.
As mentioned before, we note  that $T_{j-1}^{S*}$ only influences $\tau_j^*$ through $P(\omega | T_{j-1}^{S*} M_{j-1})$.
Besides this, the values $\tau_1^*,\dots,\tau_{j-1}^*$ are not used explicitly in determining $\tau_j^*$.  This type of optimization, in which the figure of merit in the nearest future step is maximized\cite{fischer2000quantum-state,ferrie2012adaptive,huszar2012adaptive} is also called `greedy' strategy.

\begin{figure}
    \centering
    \includegraphics{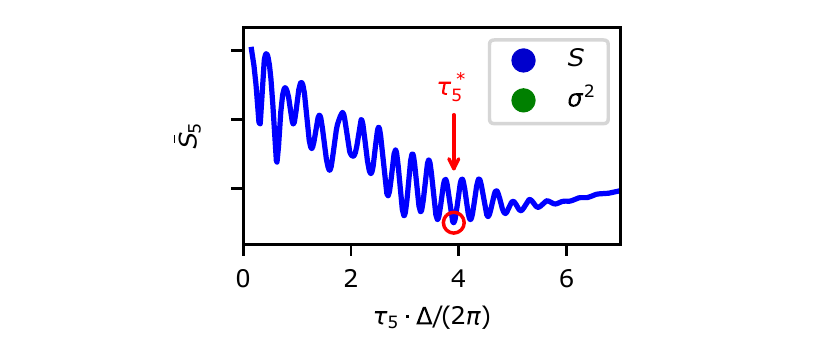}
    \caption{Illustration of the choice of $\tau_5^*$ with Eq.~\eqref{eq:S_j}.
    }
    \label{fig:oscillating_S}
\end{figure}

\begin{figure}
    \centering
    \includegraphics{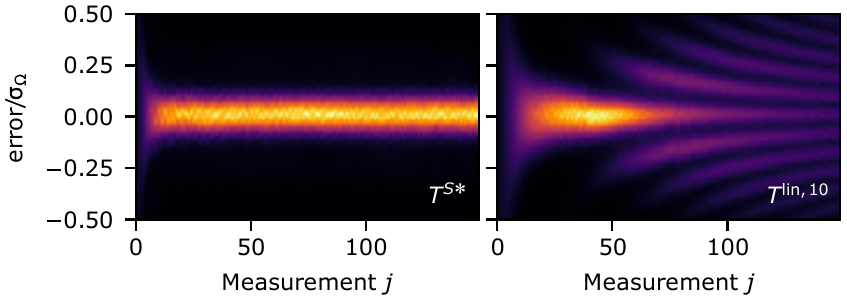}
    \caption{Mean error distributions (with respect to the true value of $\omega$) as a function of $j$.}
    \label{fig:averaged_bayesians}
\end{figure}

Fig.~\ref{fig:oscillating_S} illustrates the calculation of $\tau_5^*$ with Eq.~\eqref{eq:S_j}.
An analytical calculation is not possible, but here are two qualitative remarks.
First, $\bar{S}_j$ has a global minimum.
For too short $\tau_j$, the Bayesian update yields a smeared posterior.
On the contrary, for $\tau_j$ too long, the posterior shows multiple narrow peaks.\cite{danilin2018quantum-enhanced}
Fig.~\ref{fig:averaged_bayesians} clearly reflects these limiting cases.
The optimal $\tau_j^*$ lies in between.
For much greater $\tau_j$ values, a high damping $\psi(\tau)$ in Eq.~\eqref{eq:closed} makes the measurement less informative and eventually the entropy saturates to a constant value.
Second, $\bar{S}_j$ shows many local minima and maxima.
They correspond to the oscillations of $\cos\int_{t_i}^{t_i+\tau} dt\, \Omega(t)$ in Eq.~\eqref{eq:Pmt}.
We interpret them as follows.
A projective measurement is most informative when the two binary outcomes $m = \pm 1$ are equally probable and it is least informative when there is only one possible outcome.
Changing $\tau_j$ alternates between these cases and makes the entropy oscillate.

Let us evaluate the accuracy of our protocols with feedback.
This analysis includes
(i) $\langle S_T \rangle$ in Fig.~\ref{fig:comparison_S};
(ii) the inset of Fig.~\ref{fig:comparison_S}, with the so-called \emph{odds} or ratios of likelihoods $\exp(\langle S_{T^r} \rangle - \langle S_{T^{S*}} \rangle)$, $T^r \in \{T^S, T^{\sigma*}, T^\sigma, T^{\rm{saw}}, T^{\rm{lin},1}, \dots, T^{\rm{lin},K}\}$ ($T^S$ and $T^\sigma$ defined below);
and (iii) Fig.~\ref{fig:averaged_bayesians}, with the average error at $t = t_N$ with respect to the true value.
While (i) and (ii) mainly concern the entropy and likelihood, related through Eq.~\eqref{eq:theorem}, (iii) is linked to the variance.
We find the following:

\begin{figure}
    \centering
    \includegraphics{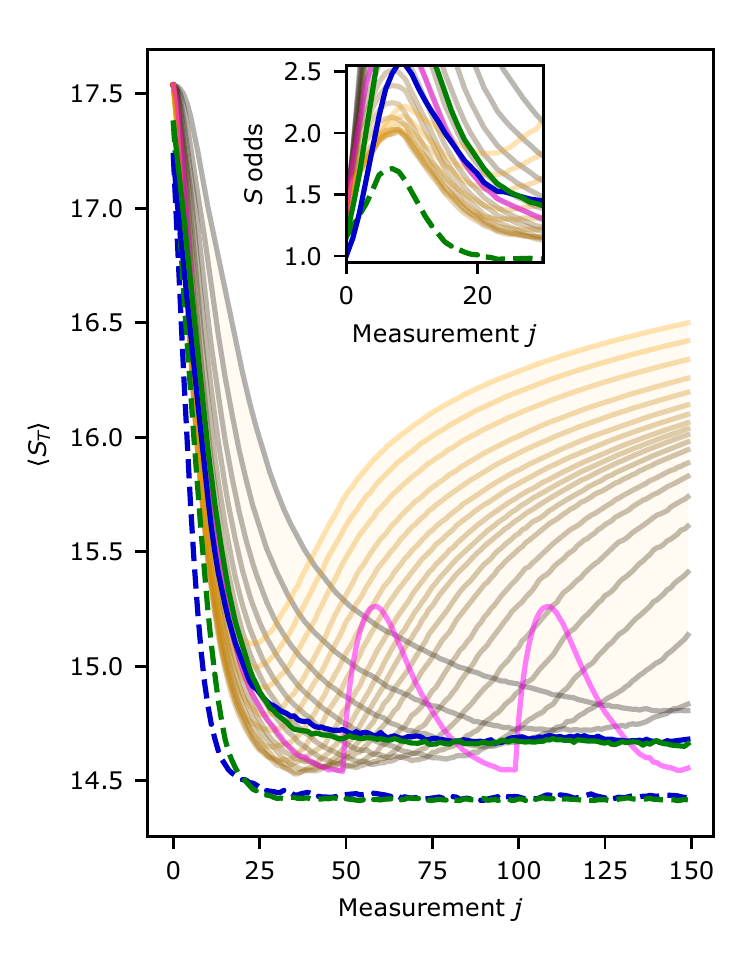}
    \caption{Comparison between protocols $T^{S*}$, $T^S$, $T^{\sigma*}$, $T^\sigma$, $T^{\rm{lin},k}$, and $T^{\rm{saw}}$.
    $\langle S_T \rangle$ is the mean entropy of the posterior $P(\omega | T_j M_j)$ as a function of $j$ for all protocols.
    The average is taken over a horizontal plane in the configuration space of Fig.~\ref{fig:configuration_space}b.
    Dashed (solid) lines correspond to protocols with (without) feedback; blue (green) to entropy (variance) optimization; the gradient from gray to yellow to $T^{\rm{lin},k}$ with increasing slope $\alpha_k$; and magenta to $T^{\rm{saw}}$.
    The same color code applies to Fig.~\ref{fig:meas_patts}.
    The faint yellow shadow covers the region swept by linear protocols as a guide for the eye.
    The inset plots the odds of $T^{S*}$ versus the other protocols, as defined and discussed in the main text.}
    \label{fig:comparison_S}
\end{figure}

First, Fig.~\ref{fig:comparison_S} makes clear that the protocols with feedback $T^{S*}$ and $T^{
\sigma*}$ beat all other protocols.
They reach a lower value of $\langle S_T \rangle$ after fewer measurements and indefinitely sustain it.
$T^{\rm{lin},k}$ eventually give worse results.
They also lead to multiple peaks in Fig.~\ref{fig:averaged_bayesians}.
The same happens with $T^{\rm{saw}}$, as Sec.~\ref{sec:nofeedback} discusses.
In contrast, protocols with feedback give a single narrow peak with less variance.

Second, as Sec.~\ref{sec:mle} anticipated, $T^{S*}$ and $T^{\sigma*}$ are not equivalent.
$T^{S*}$ saturates $\langle S_T \rangle$ faster, see the inset in Fig.~\ref{fig:comparison_S}.
Therefore, it is the best option for \emph{localization}.
Using a figure (not shown) analogous to Fig.~\ref{fig:comparison_S} but plotting variance instead of entropy, we have checked that $T^{\sigma*}$ gives a slightly lower mean variance.
In any case, since both $T^{S*}$ and $T^{
\sigma*}$ eventually saturate to the same $\langle S_T \rangle$, we conclude that their performance is comparable.

\begin{figure}
    \centering
    \includegraphics{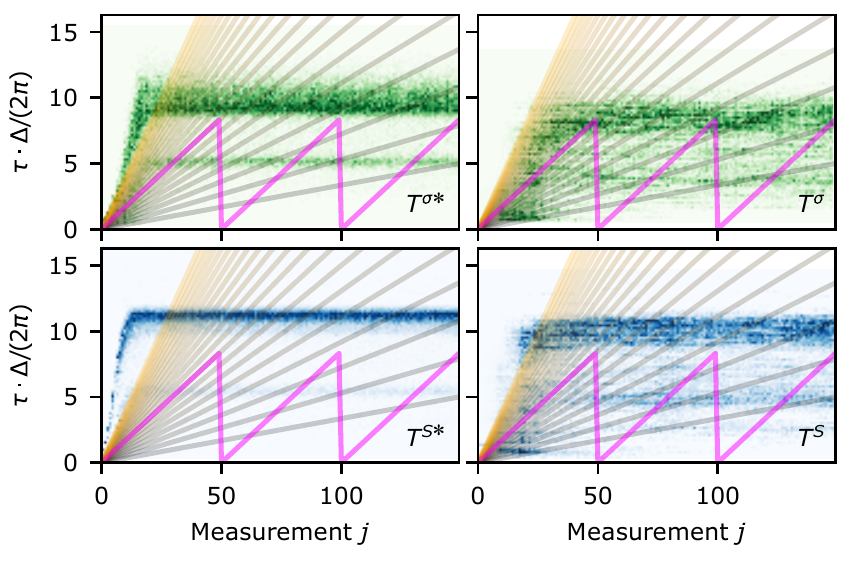}
    \caption{Distributions of $\tau_j^*$ values for protocols with feedback (the left column) and the converged distributions $\tilde{P}(\tau_j)$ for protocols without feedback (the right column), with $T^{\rm{lin},k}$ and $T^{\rm{saw}}$ overlaid.
    As labeled with the text symbols, in the upper row the figure of merit is variance, in the lower it is the entropy.
The color legend is consistent with Fig.~\ref{fig:comparison_S}.}
    \label{fig:meas_patts}
\end{figure}

The core of our protocols are the distributions of the evolution times $\tau_j$. 
For protocols with feedback, they are plotted in the left column of Fig.~\ref{fig:meas_patts}.
We highlight the following remarks.
First of all, although the choice of $\tau_j$ is memoryless, the distributions show a definite structure as a function of $j$.
They begin with a rapid linear ascent which saturates to a plateau.
These two regions match the convergence to and maintenance of the minimum $\langle S_T \rangle$ in Fig.~\ref{fig:comparison_S}, respectively.
Secondly, there is a faint replica of this structure, roughly scaled to half in the $\tau$ axis.
Most likely, its role is to quenche the multiple peaks in $P(\omega | T_jM_j)$ when they appear.
Finally, despite the quantitative differences, $T^{S*}$ and $T^{\sigma*}$ share these features.

We emphasize the robustness of our protocols.
Without any phenomenological parameters, their performance is excellent.
They work for arbitrary diffusion speed and automatically handle perturbations like experimental errors: After all, they optimize the figure of merit regardless of the system details (here encompassing the Hamiltonian, the measurement, and the noise descriptions).
Consequently, our protocols can be directly applied to very diverse systems.
This is in contrast to previous works.
For example, Refs.~\onlinecite{dinani2019bayesian,cappellaro2012spin-bath} modify Kitaev's algorithm by repeating each $\tau_j$.
The number of repetitions is the parameter tuned numerically.
This approach seeks optimality with a single parameter, but why not use more?
In order to sustain the maximum precision, when should one restart that protocol?
How to modify it when experimental errors happen?
Our protocols do not raise this kind of questions.

Despite their good performance, protocols with feedback require costly calculations.
The diffusion to compute the curve $\bar{S}_j$ is the most limiting.
For this reason, they cannot be applied to experiments with fast measurement rates.
Without feedback, one can trade this cost for a reasonable compromise in precision.
We analyze to what extent this is the case in the next subsection, while keeping the robustness and avoiding ad hoc choices of the interaction times.

\subsection{Protocols without feedback}
\label{sec:nofeedback}

Protocols without feedback set $T = \{\tau_1, \dots, \tau_N\}$ before the experiment starts.
Examples in the literature are $T^{\rm{saw}}$,\cite{delbecq2016quantum} defined earlier as $\tau_j = \alpha_m(j \mod n)$, or Kitaev's protocol: $\tau_j = 2^{N-j} \tau_0$, $j = 1, \dots, N$.\cite{dinani2019bayesian}
In our protocols without feedback, instead of giving a deterministic expression for $\tau_j$, we take a different and novel approach.
We generate each $\tau_j$ from a distribution $\tilde{P}(\tau_j)$ (notice the dependence of $\tilde{P}$ on $j$, according to the usual notation
\footnote{
The probability distributions of two different variables $x$ and $y$ are usually represented by $P(x)$ and $P(y)$, although $P(x = a)$ does not necessarily equal $P(y = a)$.
The same applies to our distributions $\tilde{P}(\tau_j)$, $j = 1, \dots, N$.
}
in probability theory).
In other words, we pick $\tau_j$ randomly with probability $\tilde{P}(\tau_j)$ before the experiment starts.
Thus the label \emph{probabilistic} from now on.
But what is the $\tilde{P}(\tau_j)$ that makes the protocol optimal?
We present the self-consistent method to construct it.
As we discuss later on, self-consistency guarantees optimality.
In fact, the construction of the set of distributions $\tilde{P}(\tau_j)$ means adapting to the particular features of the system.

\begin{figure*}
    \centering
    \includegraphics{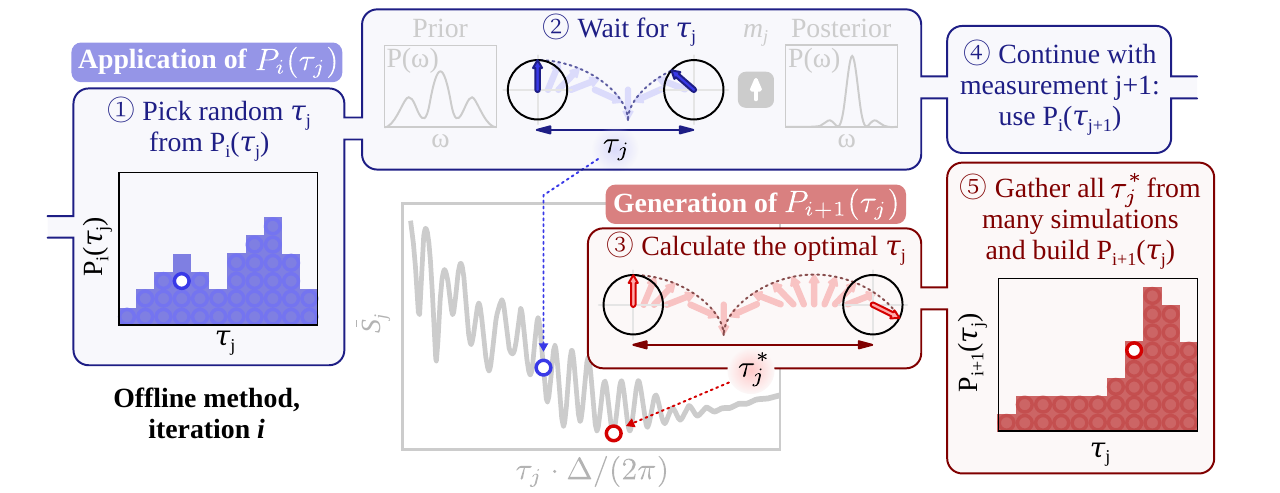}
    \caption{
    Outline of the self-consistent method to obtain $\tilde{P}(\tau_j)$, depicting iteration $i$ and measurement $j$.
    Blue corresponds to the application of the protocol generated with $P_i(\tau_j)$.
    Red only gathers the information to construct $P_{i+1}(\tau_j)$.
    The plot of $\bar{S}_j$ shows that the red $\tau_j^*$ is better than the blue $\tau_j$.
    This information will be used in the next iteration $i+1$.
}
    \label{fig:nofeedback}
\end{figure*}

An outline of our \emph{probabilistic self-consistent protocol} appears in Fig.~\ref{fig:nofeedback}.
The method proceeds iteratively.
In the iteration $i$, pick the values $\{\tau_1, \dots, \tau_N\}$ randomly with probabilities given by the initial distributions $\{P_i(\tau_1), \dots, P_i(\tau_N)\}$, respectively (for $i=1$, take $\tau_j$ arbitrarily).
Next, while an experiment is running (or more often, being simulated) with that protocol (see blue in Fig.~\ref{fig:nofeedback}), compute also the optimal $\tau_j^*$ for each $j$ with Eq.~\eqref{eq:S_j} (red in Fig.~\ref{fig:nofeedback}).
Once enough data have been collected, construct the distributions (by histograms or other parametrization) $\{P_{i+1}(\tau_1), \dots, P_{i+1}(\tau_N)\}$ out of those optimal $\tau_j^*$.
These distributions feed the next iteration, $i+1$.
Repeat the process until approaching the limit 
\begin{align}
    \tilde{P}(\tau_j) \equiv \lim_{i\to\infty} P_i(\tau_j)
    \,.
    \label{eq:P_tilde}
\end{align}
As we show below, only a few iterations suffice

The convergence to this limit is critical to apply our method.
Let us prove by induction its existence and uniqueness.
Assume $\tilde{P}(\tau_k) = \lim_{i\to\infty} P_i(\tau_k)$ exists and is unique for $k = 1,\dots,j$, and let $T_j = \{\tau_1, \dots, \tau_j\}$ be a protocol generated by $\tilde{P}(\tau_k)$.
$(M_j, T_j)$ uniquely determines $\tau_{j+1}^*$ through Eq.~\eqref{eq:S_j}, and thus $\tilde{P}(\tau_{j+1})$ is the distribution of $\tau_{j+1}^*$ after sampling the whole subspace $(M_j, T_j)$.
Moreover, $\tilde{P}(\tau_1) = \lim_{i\to\infty} P_i(\tau_1) = \delta(\tau_1-\tau_1^*)$ trivially exists because $\tau_1^*$ only depends on the prior $P(\omega)$.
This completes the proof.
Note that this reasoning does not impose anything on the protocol $T$.
For this reason, the method of Eq.~\eqref{eq:P_tilde} can optimize general protocols without feedback as we discuss in the end of this section.

\begin{figure}
    \centering
    \includegraphics{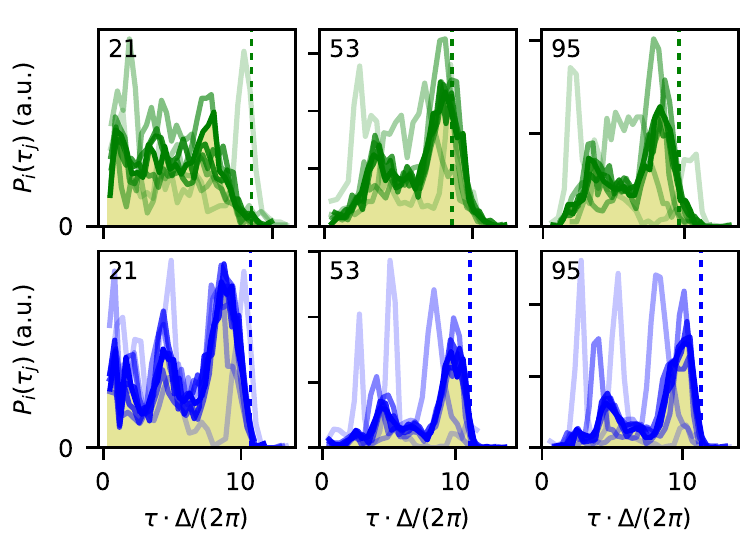}
    \caption{$\tau_j$ distributions for the measurement $j$ displayed on the top left corners.
        The results for the iterations $i = 1,\dots,8$ are shown with increasing opacity, and the seed for $\tau_j$ with a dotted line.
        The limit distributions $\tilde{P}(\tau_j)$, filled with a yellow shadow as a guide for the eye, are the cuts of the density plots in Fig.~\ref{fig:meas_patts}.
        Blue (green) corresponds again to the variance (entropy).}
    \label{fig:selfcon_tau_evolution}
\end{figure}

Our method can only be applied in practice if $P_i(\tau_j)$ converges to $\tilde{P}(\tau_j)$ fast.
For our example, this is proved in Fig.~\ref{fig:selfcon_tau_evolution}:
Convergence is reached after less than 8 iterations.
Additionally, an adjustment on the calculation of $\tilde{P}(\tau_j)$ allows us to speed up the method.
It consists of applying the self-consistent method described above but only for the distributions $\{\, \tilde{P}(\tau_j) \mid j \in r\mathbb{N}, \ 1 < r < N \,\}$.
We take $r = 15$.
The remaining distributions $\{\, \tilde{P}(\tau_k) \mid k \notin r\mathbb{N} \,\}$ are linearly interpolated.

With the method of Eq.~\eqref{eq:P_tilde}, we generate the protocols $T^S$ and $T^\sigma$.
Let us analyze what precision they achieve.
Fig.~\ref{fig:comparison_S} shows that they perform similarly to $T^{S*}$ and $T^{\sigma*}$, but there are two main differences.
First, as expected, the minimum $\langle S_T \rangle$ from $T^S$ and $T^\sigma$ is slightly above the minimum from $T^{S*}$ and $T^{\sigma*}$.
The odds quantify the difference in terms of likelihood, stabilizing at $\sim 1.5$.
Second and more strikingly, several linear protocols surpass $T^S$ and $T^\sigma$ during the first half of the experiment.

It does not necessarily mean that linear protocols are better.
Indeed, for large $j$, $T^S$ and $T^\sigma$ eventually outperform $T^{\rm{lin},k}$ for any $k$.
This happens when $\tau_j$ in $T^{\rm{lin},k}$ becomes too large, producing multiple peaks in Fig.~\ref{fig:averaged_bayesians} (right). 
But then the question is: Can the protocols $T^{\rm{lin},k}$ be modified to sustain the minimum $\langle S_T \rangle$ they get to, and therefore beat $T^S$ and $T^\sigma$?
This is what we aim at with $T^{\rm{saw}}$.
It restarts a linear protocol when it reaches the minimum $\langle S_T \rangle$, expecting to maintain that value from then on.
However, Fig.~\ref{fig:comparison_S} shows that rather than keeping $\langle S_T \rangle$ constant, $T^{\rm{saw}}$ makes it oscillate.
$T^{\rm{saw}}$ periodically recovers the minimum $\langle S_T \rangle$, but on average $T^S$ and $T^\sigma$ perform better.
We expect the same behavior for any other protocol without feedback.
We conjecture that the protocol generated by Eq.~\eqref{eq:P_tilde} is the ultimate protocol without feedback to sustain the maximum precision indefinitely.

Playing the central role, we now examine the limit distributions $\tilde{P}(\tau_j)$, plotted in the right column of Fig.~\ref{fig:meas_patts}.
As expected, they share the main qualitative features with their left-column counterparts, discussed in Sec.~\ref{sec:feedback}.
Slight differences are that $\tilde{P}(\tau_j)$ are more smeared, and that the replicated plateaus display more weight.
Once again, the important point is that optimality does not depend on imposed features or heuristic parameters.
The true value of self consistency is this robust and automatic tuning.
Self consistency straightforwardly handles measurement errors or wide variations of the noise-dynamics parameters and leaves the construction of the protocol expressed in Eq.~\eqref{eq:P_tilde} intact.

We conclude by pointing out the broad applicability of our probabilistic self-consistent protocol.
Not only can it optimize the memoryless protocols we focused on, but it would apply to more general cases.
This is a direct consequence of the proof of existence and uniqueness we gave after Eq.~\eqref{eq:P_tilde}.
For example, we can use self-consistency to improve any protocol without feedback in the literature.
Let the protocol be originally $\tau_j = \tau(j)$, with $\tau(j)$ a certain  function, and let us improve it to $\tau_j^*$ within the constraint $\tau_j^* \in [\tau(j)-L(j), \tau(j)+L(j)]$.
$L(j)$ must be large enough for this interval to contain at least one local minimum in the figure of merit, see Fig.~\ref{fig:oscillating_S}.
We choose the optimal $\tau_j^*$ as one of those minima.
The rest of our method stays the same, and so we generate a protocol without feedback.
Remarkably, this combines self-consistency with the \emph{memory} provided by the function $\tau(j)$.
Modifications like this would respond to self-consistency rather than to the heuristic parameters used in the literature.\cite{cappellaro2012spin-bath,dinani2019bayesian,higgins2009demonstrating}
We leave these optimizations as a continuation to this work.

In sum, this section proves that our self-consistent method (i) yields good precision, only slightly less than protocols with feedback; (ii) keeps that precision stable in time, outperforming other protocols without feedback and preserving more coherence; and (iii) can improve any other protocol without feedback.
For these reasons, we propose our self-consistent probabilistic protocol as a robust way to optimize generic protocols without feedback.
This is the main result of this article.

\section{Discussion}

Here we extend the discussion started in the Introduction.
In this part, we discuss connections to existing results aiming at an expert in the field.
This section is not required to understand our method but might be helpful to make the differences to existing methods more transparent.

\label{sec:discussion}

\subsection{Relevance of the dynamics}

\label{sec:noise}

 We begin with a remark on  terminology.
We separate the effects called \emph{noise} in the Introduction into two different categories: \emph{dynamics} stands for the inherent instability of the estimated variable, and \emph{errors} stands for experimental imperfections, such as in the measurement, evolution (for example, the actual evolution time differs from the intended one), initialization, and so on.\cite{demkowicz-dobrzanski2009quantum}
In this list of \emph{noise}, we do not include \emph{quantum projection noise} (quantum-mechanical measurements results being probabilistic).

Many of the existing strategies start from a solution of an idealized problem, QPEA being a typical example, ignoring dynamics.
The latter effects are taken into account at the end, or somewhere within (or also nowhere in) the optimization algorithm in an intuitive way:
Typically, as a restriction on the evolution time $t$, derived from some dephasing-time scale.
However, doing so,  the optimal solution becomes subjective: Depending on how the dephasing time is defined, at which point it is taken into account, with what prefactors or weights, and so on.
In our approach, such dephasing is not an additional input parameter.
Instead, dephasing emerges from the statistical properties of the estimated variable [see Eq.~\eqref{eq:closed}].
We view such internal consistency as important\cite{macieszczak2014bayesian} and consider the nature of the dynamics as an inevitable and inseparable part of the estimation-problem formulation.
But there are more reasons.
Indeed, were the estimated variable stable, there would be no need for optimal estimation: The variable would be estimated, even if in a suboptimal way, once and for all.
To show that this issue is not just academic, we point out several practical consequences of ignoring the instability of the estimated variable.

First, estimators are routinely compared from their behavior at large $N$.
In numerics a large $N$ is needed to establish the scaling reliably, while taking the limit $N \to \infty$ has an obvious appeal for analytics. \cite{braunstein1992quantum} But the behavior at $N \to \infty$, implying $\textrm{var}(\phi) \to 0$, is, strictly speaking, irrelevant: The limit can not be reached, as the variance is limited from below due to the dynamics in any realistic scenario.
In Fig.~\ref{fig:comparison_S} the decrease of the variance is stopped by the dynamics, for $N$ as small as about $20$.
While one expects that the estimator which is best at $\textrm{var}(\phi)\to 0$ might also be good for finite variances, one cannot assume that it is best there as well.

Second, ignoring dynamics might look legitimate in the initial stage, where the uncertainty due to the prior dominates the dynamics.
We call this regime \emph{localization}, while the one where the variance saturates is \emph{tracking}.\footnote{The name \emph{tracking} is well agreed on,\cite{doucet2001sequential} although Ref.~\onlinecite{higgins2009demonstrating} uses \emph{sensing} instead.
We are not aware of a unified name for the \emph{localization} regime.
\emph{Detection} as used in image reconstruction is related, but not the same.}
In the latter case, there is a balance between the entropy influx due to dynamics and its erasure through measurements.
A view of tracking as such a balance appeared in several works.\cite{bonato2017adaptive,berry2002adaptive,ferrie2013how} Obviously, the tracking regime can not arise at all if dynamics is ignored.
In our case the balance is reached automatically; both localization and tracking regimes are addressed optimally within a single algorithm.

Third, while technical, the following point is worth mentioning: If the dynamics is ignored, then the entropy of the next posterior can no longer be used as a figure of merit.
Indeed, independent on the prior, without dynamics the posterior entropy is minimal for $t_n \to \infty$, an unphysical result. 
Taking dynamics into account resolves the issue, and a well-defined minimum in Fig.~\ref{fig:oscillating_S} appears.
The upturn in the curve for long times is due to dynamics only.
We note that this issue is an artifact of replacing the global optimization by a local one.
Alternatively, the issue can be resolved by using entropy gain penalized by the interaction time, as was done in Refs.~\onlinecite{ruster2017entanglement-based,berry2009how,mitchell2005metrology}.

Finally, we note that abandoning the $N\to\infty$ scaling as the way to assess estimation schemes, there is no agreed way to judge how our two estimation protocols, with and without feedback, differ. This could be qualitatively, analogous to different scaling in $N$, or only quantitatively, with the same scaling in $N$ but with different prefactors.
We leave devising a suitable comparison measure (perhaps based on the information metric, see Footnote \onlinecite{FootnoteFOM}) as a task for the future.

\subsection{Are quantum resources needed to achieve Heisenberg scaling?}

\emph{Quantum metrology} holds the promise that ``quantum effects enable an increase in precision when estimating a parameter''.\cite{giovannetti2006quantum} That is, using quantum resources enables overcoming limits imposed by classical physics.
A paradigmatic case, related to our topic, is overcoming the SQL and reaching the ultimate Heisenberg limit.
The works of Caves \cite{caves1980quantum-mechanical,caves1981quantum-mechanical} were instrumental for recognizing the possibility of it.
Importantly, it was soon realized that the decoherence will put a strong restriction on what is practically achievable.
The original observation that it nullifies any gain in a typical scenario with NOON-states,\cite{huelga1997improvement} was later generalized into a formal inequality, first for limit cases\cite{knysh2011scaling,kolodynski2010phase} and then in general\cite{escher2011general}. 

Nevertheless, putting the decoherence aside, it is worth looking briefly at  the role of quantum resources in phase estimation and their historical development.
Obviously, the QPEA, being a quantum computing circuit, is essentially quantum.\cite{cleve1998quantum} However, Ref.~\onlinecite{griffiths1996semiclassical} showed that the entanglement can be traded for classical feedback. 
In line with that, Refs.~\onlinecite{higgins2007entanglement-free,berry2001optimal,berry2000optimal} conclude that with feedback the estimation reaching Heisenberg scaling is indeed possible.
Finally, however, the link through Ref.~\onlinecite{griffiths1996semiclassical} seems to be removed, once it was found that even feedback is not necessary,\cite{higgins2009demonstrating} and that  methods without feedback can also reach Heisenberg scaling.\cite{bonato2016optimized}
The object on which we estimate (a spin $1/2$) is certainly quantum.
However, as there is no entanglement needed (not even effectively through feedback), the protocol could (and should) be called ``classical''.

\section{Conclusions}
\label{sec:conclusions}

We have analyzed the optimization of frequency estimation for a two-level system in the presence of arbitrarily fast dephasing sources.
We considered protocols with and without feedback and two archetypal figures of merit, the squared error and entropy, in turn related to maximum likelihood.

Despite being memoryless, the protocols with feedback outperform all others we tried. 
Moreover, they indefinitely sustain the maximum precision attained.
Most remarkably, we have designed a protocol without feedback that performs almost identically and (i) can also optimize generic protocols for any figure of merit and does not require any heuristic input; (ii) can be applied to experiments with arbitrarily fast measurement rates; (iii) is robust under general circumstances, including measurement errors; and (iv) is numerically feasible, constructed after a few self-consistent iterations.

In summary, we have designed and tested a versatile protocol that can significantly improve the precision in parameter estimation.
Among other prospects, it might allow for increased coherence times of solid-state qubits.

\section{Acknowledgments}
\'A. Guti\'errez-Rubio acknowledges the project CREST JST (JPMJCR1675) through the Japanese Science and Technology Agency.
This work was supported by  the Swiss National Science Foundation and NCCR SPIN.

\appendix

\section{Generalizations of the system}
\label{app:hamiltonian}

In this appendix, we give some guidelines to generalize our numerical implementation of the optimal protocol construction.
Namely, we analyze how the design changes (i) for a general two-level Hamiltonian and (ii) under measurement errors.
Finally, we briefly point out further possible extensions.

We focus first on a general two-level system.
Let $H(t) = H_0 + H'(t)$, with
\begin{align}
    H_0 = (\hbar \Delta/2) \hat{n} \cdot \vec{\sigma} \,,
    \quad
    H'(t) = [\hbar \Gamma(t) / 2] \sigma_x \,,
    \label{eq:h_general}
\end{align}
$\Delta > 0$ constant, $\hat{n} = n_x\hat{x} + n_y\hat{y} + n_z\hat{z}$ a unit vector, and $\Gamma(t)$ a stochastic variable.
We assume $\langle\Gamma(t)\rangle = 0$.  The eigenstates of $\sigma_z$ are denoted by $\ket{\uparrow}$ and $\ket{\downarrow}$ and define our space for projective measurements.
Further, we define the frequency $\Omega(t) = |\Delta \hat{n} + \Gamma(t)\hat{x}|$, where
$\Gamma(t)$ is the source of dephasing with respect to the average $\Delta = \langle \Omega(t) \rangle$.
The goal is to estimate $\Gamma(t)$ or, equivalently, $\Omega(t)$.
One can map this model to a variety of systems where dephasing has different origins.
For example, with $\hat{n} = \hat{x}$, $H(t)$ maps to the double dot discussed in Sec.~\ref{sec:system} or to holes with Ising-like interactions.\cite{fischer2008spin}

Importantly, almost all the expressions and protocols in the main text remain valid for the general Hamiltonain $H(t)$ of Eq.~\eqref{eq:h_general}.
But there is one important change.
Now, the commutator of the Hamiltonian with itself at different times is different from zero:
\begin{align}
    [H(t), H(t')] = 2i \Delta [\Gamma(t)-\Gamma(t')]
    (n_z \sigma_y + n_y \sigma_z) \,.
    \notag
\end{align}
Therefore, in Eq.~\eqref{eq:Pmw} we must insert
\begin{align}
    P(m | \Omega, \tau) = |\bra{m}
    T \exp
    \left[
        -\frac{i}{\hbar} \int_{t_i}^{t_i+\tau}dt \, H(t)
    \right]
    \ket{\uparrow}|^2
    \notag
\end{align}
and we cannot drop the time-ordering operator $T$.
Consequently, there is no simple expression for $P(m | \Omega, \tau)$ like Eq.~\eqref{eq:Pmt}, and no closed expression for $P(m | \omega, \tau)$ like Eq.~\eqref{eq:closed}.
In sum, Bayesian update is now a difficult task.

A workaround is to approximate $\Gamma(t) \simeq \Gamma(t_i)$ for $t \in [t_i, t_i+\tau]$.
As we discuss in the main text, this approximation is valid only for negligible diffusion during interaction times $\tau$.
In this case,
\begin{align}
    P(m | \omega, \tau) \simeq
    \frac{1}{2}
    \left[
        1 + m
        \left(
            \cos^2\beta + \sin^2\beta\cos(\omega \tau)
        \right)
    \right] \,,
    \label{eq:beta}
\end{align}
with $\omega = \Omega(t_i)$ and $\cos\beta = \hat{n}\cdot\hat{z}$.
Always within the regime of slow diffusion, this expression extends our protocols to general two-level systems.
One should use it instead of Eq.~\eqref{eq:closed}, but the rest of our analysis remains the same.

Now, we extend our study with measurement errors.
Let $\eta_\uparrow$ ($\eta_\downarrow$) be the probability to make an error when measuring $\ket{\uparrow}$ ($\ket{\downarrow}$).
Define $\mu = \eta_\downarrow - \eta_\uparrow$ and $\nu = 1 - \eta_\downarrow - \eta_\uparrow$.
In the main text, Eq.~\eqref{eq:Pmt} turns into
\begin{align}
    P (m | \Omega, \tau)
    = \frac{1}{2}
    \left[
        1 + m
            \left(
                \mu + \nu\cos
                \int_{t_i}^{t_i+\tau} dt\, \Omega(t)
            \right)
    \right] \,,
    \notag
\end{align}
and Eq.~\eqref{eq:closed} changes accordingly.
For the general Hamiltonian of Eq.~\eqref{eq:h_general}, Eq.~\eqref{eq:beta} takes the form
\begin{align}
    P(m | \omega, \tau) \simeq
    \frac{1}{2}
    \left\{
        1 + m
        \left[
            \mu + \nu
                (\cos^2\beta + \sin^2\beta\cos(\omega \tau)
        \right]
    \right\} \,.
    \notag
\end{align}
This expression shows the equivalence between measurement errors on the one hand, and a nonzero $\cos\beta = \hat{n} \cdot \hat{z}$ (coming from $\sigma_y$ and $\sigma_z$ terms in the Hamiltonian) on the other.
Either of them separately, or both together, yield $r + s\cos(\omega t)$ inside the square brackets, with $r + s = 1$.
Their effect is to make the Bayesian update less efficient in narrowing the prior.
Technically, a finite value of $r=1-s$ prevents $P(m | \omega, \tau)$ from reaching zero values.
Such zeros are desirable because they discard frequencies when applying Bayes' rule, see Eq.~\eqref{eq:Bayes}.

One can also think of further generalizations.
For example, the perturbation in Eq.~\eqref{eq:h_general} could be $H'(t) = (\hbar/2)\sum_i \Gamma_i(t)\sigma_i$, with $i$ running over $\{x, y, z\}$.
This would require us to estimate three stochastic variables instead of one.
The Bayesian formalism, although more involved, would stay the same.
At last, realistic noise often has a more complicated kernel than Eq.~\eqref{eq:kernel}.
Similarly, the diffusion of probability distributions through Eq.~\eqref{eq:diffusion} does not cover the most general case.
Using ARMA models seems a viable option to simulate such noise.
We leave these studies as a possible application of our main idea.

\section{Proof of Eq.~\eqref{eq:closed}}
\label{app:proof_closed}

To prove Eq.~\eqref{eq:closed}, we insert Eq.~\eqref{eq:Pmt} into Eq.~\eqref{eq:Pmw}.
Then, substituting $\cos x = (e^{ix} + e^{-ix})/2$ and using $\int \mathcal{D}\Omega\, P(\Omega) = 1$, we can obtain the result from the value of
\begin{align}
    E \equiv
    \int \mathcal{D}\Omega \, P(\Omega)
    \exp\left(
        i\int_{t_i}^{t_i+\tau} dt\, \Omega(t)
        \right) \,.
    \notag
\end{align}
The remaining part of this section explains how to calculate this expression.

Within the time interval $[t_i, t_i+\tau]$, consider the points $\theta_\alpha = t_i + \alpha \, \delta t$, with $\delta t = \tau/N$ and $\alpha = 0, \dots, N$.
We will take the limit $N \to \infty$ in the end.
Defining $\omega_\alpha = \Omega(\theta_\alpha)$, the functional integral in $\Omega$ can be split as:
\begin{align}
    & \int \mathcal{D}\Omega =
    \prod_{\alpha=1}^N \int d\omega_\alpha \,, \notag\\
    & P(\Omega) = \prod_{\alpha=0}^{N-1}
    K(\omega_\alpha - \Delta, \omega_{\alpha+1} - \Delta, \delta t)
    \,, \notag\\
    &
    \exp\left(
        i\int_{t_i}^{t_i+\tau} dt\, \Omega(t)
        \right)
        = \sum_{\alpha=0}^N e^{i \omega_\alpha \delta t}
    \,. \notag
\end{align}
For the second expression, notice that the kernel $K(\omega_\alpha-\Delta, \omega_{\alpha+1} - \Delta, \delta t)$ is the probability that the frequency changes from $\omega_\alpha$ to $\omega_{\alpha+1}$ after a time $\delta t$.
Recall from the main text that $\Delta$ is the mean value of $\Omega(t)$.

$E$ can then be calculated, after the change of variables $\omega_\alpha \to \omega_\alpha + \Delta$, by performing the integrals in $\omega_N, \omega_{N-1}, \dots, \omega_1$ in that order.
That only requires the repeated use of the expression
\begin{align}
    \int_{-\infty}^{\infty} d & \omega_\beta \,
    K(\omega_\alpha, \omega_\beta, \delta t)
    e^{i F \omega_\beta \delta t} = \notag\\
    & \exp
    \left\{
        F \, \delta t \, e^{-\delta t/\kappa}
        \left[
            i\omega_\alpha
            -F \, \delta t \, \sigma_\Omega^2
            \sinh(\delta t/\kappa)
        \right]
    \right\} \,, \notag
\end{align}
with $F$ any constant independent on $\omega_\alpha, \omega_\beta$.
Finally, we take the limit $N\to \infty$, or equivalently expand to lowest order in $\delta t$.
The algebra, lengthy but straightforward, leads to
\begin{align}
    E & = \exp
    \left\{
        i\left[
        \tau \Delta + \kappa (\omega_0-\Delta)
        (1-e^{-\tau/\kappa})
        \right]
    \right\} \times \notag\\
    & \exp
    \left\{
        -\sigma_\Omega^2 \kappa
        \left[
            \tau + \frac{\kappa}{2}
            (1-e^{-\tau/\kappa})
            (-3+e^{-\tau/\kappa})
        \right]
    \right\} \,.
    \notag
\end{align}
This expression immediately yields Eq.~\eqref{eq:closed}.

\section{Proof of Eq.~\eqref{eq:theorem}}
\label{app:proof}

Log likelihood is a relevant quantity encountered in estimation and hypothesis testing.
In this appendix, we discuss its optimization on general grounds first.
Afterwards we prove Eq.~\eqref{eq:theorem}.

Let $\mathcal{H} = \{H_1, \dots, H_k\}$ be a set of hypotheses or models that presumably govern a given phenomenon, and $D = \{x_1, \dots, x_n\}$ contain all the data we have about it.
According to maximum-likelihood estimation, the hypothesis $H_j$ that best fits $D$ is the one maximizing $P(H_j | D)$.
Remarkably, this ranks the elements in $\mathcal{H}$ according \emph{only} to the available evidence $D$.

Assume that no hypothesis is preferred over any other, namely take $P(H)$ constant for all $H\in \mathcal{H}$.
By Bayes' rule, i.e.,\ $P(H | D) = P(D | H) P(H) / P(D)$, the maximum-likelihood estimation of $H$ is equivalent to the optimization of $P(D | H)$.
Thus, for logically independent data, for which $P(x_1, \dots, x_N | H) = \Pi_{i=1}^{N} P(x_i | H)$, we have
\begin{align}
    \log P(H | D) =
    \sum_{i=1}^{n} \log P(x_i | H)
    + \rm{const.}
    \label{eq:log_lik}
\end{align}
We highlight two important aspects of this result.
First of all, the performance or ranking of a general hypothesis, model or method $H$ is computed by testing it with the true data values $x_i$.
And secondly, the log likelihood is an additive function of the data.

Let us now introduce the log likelihood in our context.
Consider first a single experiment: We have a given trajectory $\Omega$, a protocol $T_A$ and the measurements $M_A$.
The final frequency has the true (and unknown in a real experiment) value $\omega = \Omega(t_N)$ and our estimation of $\omega$ is given by $P(\omega | T_A M_A)$.
The log likelihood of this single experiment is $\log P \boldsymbol{(} \omega = \Omega(t_N) | T_A M_A \boldsymbol{)}$.
This value quantifies, for this experiment we are looking at, how good the estimation with $T_A$ is.
If another protocol $T_B$ had been used on the same $\Omega$, yielding the measurements $M_B$, we would have the log likelihood $\log P \boldsymbol{(} \omega = \Omega(t_N) | T_B M_B \boldsymbol{)}$.
By comparing $\log P \boldsymbol{(} \omega = \Omega(t_N) | T_A M_A \boldsymbol{)}$ and $\log P \boldsymbol{(} \omega = \Omega(t_N) | T_B M_B \boldsymbol{)}$, one could decide whether $T_A$ or $T_B$ estimates $\omega$ better for that particular experiment.
Fig.~\ref{fig:configuration_space} illustrates this trivial comparison.

The additivity of likelihoods in Eq.~\eqref{eq:log_lik} allows us to extend this reasoning to multiple experiments.
Ideally, if one could sample the whole configuration space $(\Omega, M)$ for a given protocol $T$, the total log likelihood (averaged over the number of experiments) would add up to
\begin{align}
    L_T =
    \sum_{\Omega, M}
    \PP(\Omega M | T)
    \log P\boldsymbol{(}
        \omega = \Omega(t_N) | T M
        \boldsymbol{)} \,.
    \label{eq:pre_theorem}
\end{align}
Here, $\PP(\Omega M | T)$ represents the probability to encounter $(\Omega, M)$ given $T$.
Then, in terms of maximum likelihood, $L_T$ is the figure of merit that evaluates how well the protocol $T$ estimates the frequency.

Now, let us prove that for a given protocol $T$, $L_T$ equals the mean entropy $\langle S_T \rangle$ of all the posteriors $P(\omega | T M)$.
Departing from Eq.~\eqref{eq:pre_theorem},
\begin{align}
    L_T & =
    \sum_{\Omega, M}
    \PP(\Omega M | T)
    \log P\boldsymbol{(}
        \omega = \Omega(t_N) | T M
        \boldsymbol{)}
    \notag \\
    &
    =
    \sum_{\Omega, M}
    \PP(M | T) \PP(\Omega | T M)
    \log P\boldsymbol{(}
        \omega = \Omega(t_N) | T M
        \boldsymbol{)}
    \notag \\
    &
    =
    \sum_{\Omega, M}
    \begin{aligned}
        &
        \PP(M | T)
        \PP\boldsymbol{(}
        \Omega(t_1) \cdots
        \Omega(t_{N-1}) | \Omega(t_N) T M
        \boldsymbol{)}
        \notag \\
        &
        \quad \times
        P\boldsymbol{(}
            \Omega(t_N) | T M
        \boldsymbol{)}
        \log P\boldsymbol{(}
            \omega = \Omega(t_N) | T M
            \boldsymbol{)}
    \end{aligned}
    \notag \\
    &
    =
    \sum_{\Omega(t_N), M}
    \begin{aligned}
        &
        \PP(M | T)
        \PP\boldsymbol{(}
            \Omega(t_N) | T M
        \boldsymbol{)}
        \notag
        \\
        &
        \quad \times
        \log P\boldsymbol{(}
            \omega = \Omega(t_N) | T M
            \boldsymbol{)}
    \end{aligned}
    \notag \\
    &
    =
    -\sum_{M}
    \PP(M | T)
    S[P(\omega | T M)]
    = -\langle S_T \rangle \,. \quad (q.e.d.)
    \notag
\end{align}
This result stems from (i) the application of elementary probability rules and (ii) from the assumption $\PP \boldsymbol{(} \Omega(t_N) | T M \boldsymbol{)} = P \boldsymbol{(} \omega = \Omega(t_N) | T M \boldsymbol{)}$, namely that the sampling probability $\mathcal{P}$ in the configuration space is correctly predicted by our Bayesian estimation $P(\omega | T M)$.
Condition (ii) basically relies on the agreement between the true diffusion dynamics of $\Omega(t)$ and the kernel, see Eq.~\eqref{eq:diffusion}.
We obviously simulate our experiments under this condition.
Therefore, our theorem applies to our data.
In an actual experimental setup, however, this requirement should be confirmed by a procedure external to ours.

We conclude with a remark on the practical use of our result, Eq.~\eqref{eq:theorem}.
Eq.~\eqref{eq:pre_theorem} defines the log likelihood $L_T$ in our context.
But its computation in a real experiment is impossible in that form: As we pointed out, $P\boldsymbol{(} \omega = \Omega(t_N) | T M\boldsymbol{)}$ cannot be calculated because the true frequency $\Omega(t_N)$ is unknown.
With Eq.~\eqref{eq:theorem}, however, we can tackle this problem: The knowledge of the true frequencies $\Omega(t_N)$ is not necessary if we can calculate the average entropy of the posteriors $P(\omega | TM)$, denoted by $\langle S_T \rangle$.
Our result is valuable from a theoretical point of view, mainly discussed in Sec.~\ref{sec:mle}, but it is also a practical tool to calculate log likelihoods, as we do throughout Sec.~\ref{sec:optimal}.

\bibliography{entropy_intro,entropy,footnotes}

\end{document}